\documentclass{aa}
\usepackage{graphicx}
\usepackage{amsmath}
\usepackage{txfonts}
\usepackage{hyperref}
\usepackage{xurl}
\usepackage{capt-of}
\usepackage{orcidlink}

\hypersetup{hidelinks,pdftitle={Environmental Dependence of Galaxy Properties in the DESI DR1 Bright Galaxy Survey: Star Formation, Morphology, and AGN Activity at z < 0.55}, pdfauthor={D. Kim}}
\begin{document}

\title{Environmental dependence of galaxy properties in the DESI DR1 Bright Galaxy Survey: Star formation, morphology, and AGN activity at $z < 0.55$}

\titlerunning{Environmental dependence in DESI BGS}
\authorrunning{D. Kim}

\author{D. Kim\inst{1}\orcidlink{0000-0001-5120-0158}}

\institute{Astronomy and Space Science Department, Chungnam National University, Daehak-ro 99, Yuseong-gu, Daejeon 34134, Republic of Korea\\
\email{duhokim81@gmail.com}}

\date{}

\abstract{
We examined the environmental dependence of galaxy properties using a parent sample of about 1.4 million BGS\_BRIGHT galaxies from the Dark Energy Spectroscopic Instrument Data Release 1 Bright Galaxy Survey at $0.05 < z < 0.55$.
Using the extended halo-based group catalogue, we measured how the specific star formation rate (sSFR), S\'{e}rsic index, and active galactic nucleus fraction depend on halo mass over $9\leq\log(M_*/\mathrm{M}_\odot)<12$ and $12.5\leq\log(M_h/\mathrm{M}_\odot)<16$.
The parent measurements describe the observed sample, and our main conclusions about the galaxy population are based on comparisons above the completeness limit derived for quiescent galaxies.

Within the same stellar mass and redshift bins, the median sSFR is generally lower in clusters than in the field.
At intermediate stellar masses and $z<0.25$, satellites show a lower median sSFR than centrals at fixed stellar and halo mass.
The environmental dependence of the star formation reflects differences between centrals and satellites and the changing satellite fraction.
The S\'{e}rsic index increases only weakly with halo mass at fixed stellar mass and redshift.
Among galaxies in which all four diagnostic emission lines are detected, we find no strong environmental dependence of Seyfert or low-ionisation nuclear emission-line region fractions at fixed stellar mass. They are both marginally higher at most in clusters, with the largest but most uncertain differences at the highest stellar masses.
}


\maketitle
\nolinenumbers

\section{Introduction} \label{sec:intro}

Observations have long established that galaxy properties correlate strongly with environment, from the morphology-density relation \citep{Dressler1980} to the suppression of star formation in dense regions \citep{Gomez2003, Kauffmann2004}.
Environmental quenching (the suppression of star formation in dense environments) operates through multiple physical mechanisms, including ram-pressure stripping \citep{GunnGott1972}, strangulation \citep{Larson1980}, galaxy harassment \citep{Moore1996}, and active galactic nucleus (AGN) feedback \citep{Croton2006}.

Recent observations have begun to trace environmental quenching to higher redshifts.
Using the James Webb Space Telescope (JWST), \citet{Hamadouche2025} found strong evidence of environmental quenching of low-mass galaxies out to $z \sim 2$, observing an upturn in the low-mass end of the quiescent galaxy stellar mass function.
\citet{Mao2022} showed that the quenching efficiency in clusters at $0.5 < z < 1.0$ increases with stellar mass and environmental density, consistent with a downsizing picture.
\citet{Peng2010} demonstrated that since $z \sim 0.5$, low-mass galaxies ($\log(M_*/\mathrm{M}_\odot) < 10.5$) are quenched by environmental mechanisms rather than internal processes.
At lower redshift, \citet{Wetzel2012} showed that satellite star formation rates remain largely unaffected for several gigayears after infall before quenching rapidly, suggesting that environmental processes act on long but finite timescales.

Beyond star formation, galaxy morphology, which is quantified by the S\'{e}rsic index, also responds to the environment.
Bulge-dominated high-S\'{e}rsic-index morphologies are preferentially found in dense environments \citep{Dressler1980}, and this trend persists after controlling for stellar mass \citep{Weinmann2006}.
Recent JWST observations confirmed that the S\'{e}rsic index is strongly linked to star formation activity across redshift: quiescent galaxies consistently show higher S\'{e}rsic indices than star-forming galaxies at optical and near-infrared wavelengths, with the contrast between the two populations increasing toward lower redshifts for the most massive galaxies \citep{Martorano2025}.
This structural transformation and star formation quenching therefore appear to proceed together, although the causal ordering remains debated.

The AGN activity introduces a further dimension to the environmental picture.
AGN feedback is invoked as a key mechanism for suppressing star formation in massive haloes \citep{Croton2006}, but the observed AGN fractions show a nuanced dependence on environment: optical AGN are suppressed in dense cluster environments \citep{Kauffmann2004}, and X-ray AGN are likewise suppressed in cluster centres \citep{Koulouridis2024}.
Large samples spanning a wide dynamic range in halo mass are therefore required to disentangle AGN-driven quenching from other environmental processes.

We use the Dark Energy Spectroscopic Instrument (DESI) Data Release 1 (DR1) Bright Galaxy Survey (BGS) BGS\_BRIGHT sample \citep{DESICollaboration2025} to trace how star formation, morphology, and AGN activity depend on environment for about 1.4 million galaxies at $0.05 < z < 0.55$.
The statistical power of DESI allows us to resolve these trends simultaneously as a function of stellar mass, halo mass, redshift, and central or satellite status, providing a view of environmental quenching in the low-redshift Universe.
This paper is organised as follows: Sect.~\ref{sec:data} describes the DESI DR1 data and our sample selection; Sect.~\ref{sec:methods} outlines our analysis methods; Sect.~\ref{sec:results} presents results on environmental effects on star formation, morphology, and AGN activity; Sect.~\ref{sec:discussion} discusses implications; and Sect.~\ref{sec:conclusions} summarises our conclusions.

\section{Data} \label{sec:data}

\subsection{DESI value-added catalogues}

DESI \citep{DESICollaboration2016} is conducting the largest spectroscopic survey to date.
DR1 provides spectra for approximately 13 million galaxies.

The stellar and environmental measurements come from two DESI DR1 value-added catalogues (VACs):

(1) The stellar mass and emission line VAC (\texttt{dr1\_galaxy\_stellarmass\_lineinfo\_v1.0}; \citealt{ZouHu2024})\footnote{\url{https://data.desi.lbl.gov/public/dr1/vac/dr1/stellar-mass-emline/v1.0/}} provides stellar masses (\texttt{MASS\_CG}), star formation rates (SFRs; \texttt{SFR\_CG}), S\'{e}rsic indices (\texttt{SERSIC}), and emission line fluxes (H$\alpha$, H$\beta$, [N~{\sc ii}], and [O~{\sc iii}]) with uncertainties.
The stellar masses and SFRs were derived through CIGALE spectral energy distribution (SED) fitting \citep{Boquien2019} using broadband $g$, $r$, $z$, $W1$, and $W2$ photometry from the DESI Legacy Imaging Surveys supplemented by spectrophotometry from DESI spectra, assuming a Chabrier initial mass function (IMF).
The emission line fluxes were measured by single-Gaussian fitting after continuum subtraction performed by STARLIGHT.
The S\'{e}rsic indices were derived from single-component S\'{e}rsic profile fits to Legacy Survey imaging.

(2) The group finder VAC\footnote{\url{https://data.desi.lbl.gov/public/dr1/vac/dr1/gfinder/v1.0/}} employs the extended halo-based group-finding algorithm of \citet{Yang2007, Yang2021}.
The group finder operates on the photometric Legacy Imaging Surveys DR9 catalogue (134.7 million objects with $m_z < 21$), using spectroscopic redshifts from DESI DR1 where available and photometric redshifts otherwise.
We used its halo mass (\texttt{GRP\_LOGM}) as our primary environmental metric.
The galaxies were iteratively assigned to groups whose halo masses were estimated through abundance matching, which ranks groups by total stellar mass and matches to the expected halo mass function from simulations.
This approach is less sensitive to flux-limit incompleteness than direct richness counts at higher redshifts.
The membership rankings (\texttt{RANK}) identify the central (most massive) galaxy of each group with \texttt{RANK}~$=0$ and satellites with \texttt{RANK}~$\geq 1$.

\subsection{BGS\_BRIGHT sample selection}

We only used magnitude-limited BGS\_BRIGHT (nominal $r<19.5$). The BGS\_FAINT subsample ($19.5 \leq r < 20.175$) applies an additional colour cut based on an H$\alpha$+H$\beta$ emission-line proxy \citep{Hahn2023}, which preferentially selects star-forming galaxies and excludes a substantial fraction of objects in that magnitude range. Before constructing our analysis samples, BGS\_FAINT targets form a substantial fraction of the broader BGS group catalogue at $0.4\leq z<0.5$ and have markedly higher median specific star formation rate (sSFR) than BGS\_BRIGHT. Choosing BGS\_BRIGHT avoids this additional colour cut.
A small fraction of BGS\_BRIGHT targets have VAC-derived magnitudes fainter than $r=19.5$; we retained their catalogue classification and acknowledge this mismatch with the adopted magnitude limit.

Throughout, we write $\log M_h \equiv \log(M_h/\mathrm{M}_\odot)$ and $\log M_* \equiv \log(M_*/\mathrm{M}_\odot)$. We define environmental categories based on the halo mass: field ($\log M_h < 13$), groups ($13 \leq \log M_h < 14$), and clusters ($\log M_h \geq 14$).

We constructed our sample as follows:

\begin{enumerate}
\item We selected targets labelled BGS\_BRIGHT in the DESI catalogue (bit 1 of \texttt{BGS\_TARGET}) with a nominal magnitude limit $r < 19.5$.
\item We matched the targets to the stellar mass and emission line VAC and group catalogue, retaining galaxies with valid stellar mass, SFR, and group membership.
\item We excluded reliable quasar identifications from the preferred DESI spectra using the AGN/Galaxy Classification VAC\footnote{\url{https://data.desi.lbl.gov/doc/releases/dr1/vac/agngal/}}; the adopted SED fits do not model AGN emission.
\item We required halo masses in the range $12.5 \leq \log(M_h/\mathrm{M}_\odot) < 16$.
\item We restricted the redshift range to $0.05 < z < 0.55$.
\item We kept a broader reference sample with $8<\log(M_*/\mathrm{M}_\odot)<12.5$: 1,417,953 unique galaxies for the completeness calibration.
\item We applied the common analysis mass range, $9\leq\log(M_*/\mathrm{M}_\odot)<12$: 1,415,850 parent analysis galaxies.
\item We applied the completeness limit derived for quiescent galaxies: 709,601 selected galaxies.
\end{enumerate}

The removal of repeat observations from 1\,559\,552 spectra gives the broader reference sample shown in Fig.~\ref{fig:completeness}. The parent and selected samples both include quiescent and star-forming galaxies. We analysed the two samples, show the measurements for the parent sample in Figs.~\ref{fig:distribution}--\ref{fig:phase}, and base our main conclusions on comparisons within the selected sample.

\subsubsection{Completeness calibration and catalogue checks}

Stellar mass VAC matching.
All BGS\_BRIGHT targets classified as galaxies with reliable redshifts (\texttt{ZWARN}=0) have VAC counterparts. Unmatched targets have redshift warnings or stellar or quasar classifications.

Group-catalogue matching.
The positional cross-identification between the DESI spectroscopic catalogue and the group-finder catalogue was performed with \texttt{astropy}'s \texttt{match\_coordinates\_sky}, accepting a match when the on-sky separation was $<1\farcs0$ and the redshift agreement was $|\Delta z| < 0.01$.
Before excluding quasars, we assessed spurious catalogue matches with an offset-match test, in which the DESI positions were shifted by $\pm30''$, $\pm45''$, and $\pm60''$ and re-matched: any match recovered from the deliberately displaced catalogue was considered spurious.
Under the fiducial ($<1\farcs0$ and $|\Delta z|<0.01$) criterion, the spurious fraction, averaged over the six offset realisations, is $0.017\%$ overall and rises only weakly towards clusters. Under the positional-only criterion, it remains $\lesssim 0.3\%$ in every environment.
This matching test does not measure completeness after the analysis cuts.

\begin{figure}
\centering
\includegraphics[width=\columnwidth]{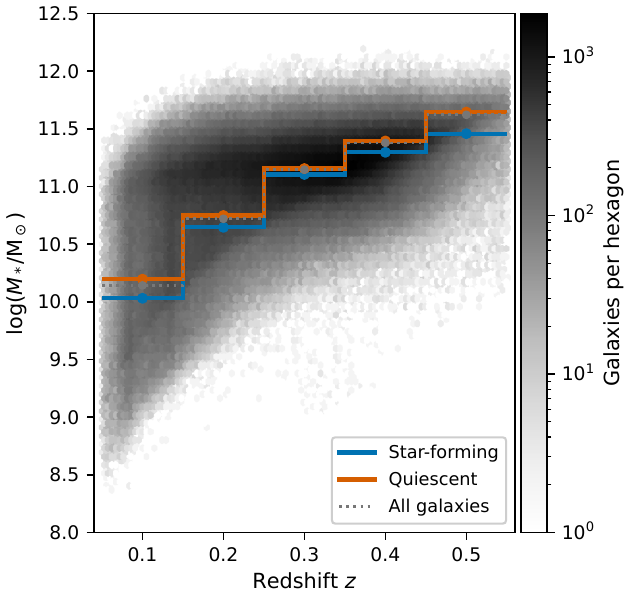}
\caption{Stellar mass completeness of the broader BGS\_BRIGHT reference sample. The grey shading shows the density in the $z$--$\log(M_*/\mathrm{M}_\odot)$ plane. The blue, orange, and grey steps give the star-forming, quiescent, and pooled 95th percentile limits \citep{Pozzetti2010} in the five $\Delta z=0.10$ analysis bins. The points mark the bin centres. Our main conclusions use galaxies above the unrounded quiescent limit.}
\label{fig:completeness}
\end{figure}

Stellar mass completeness.
Because BGS\_BRIGHT is limited by magnitude, its detectable stellar mass rises with redshift. Following \citet{Pozzetti2010}, we estimated the completeness separately for quiescent galaxies with $\log(\mathrm{sSFR}/\mathrm{yr}^{-1})<-11$ and star-forming galaxies with $\log(\mathrm{sSFR}/\mathrm{yr}^{-1})\geq-11$, using the same SED-based sSFR as in the analysis. For each galaxy, the stellar mass at the survey flux limit is $\log(M_\mathrm{lim}/\mathrm{M}_\odot)=\log(M_*/\mathrm{M}_\odot)+0.4\,(r-r_\mathrm{lim})$, with $r_\mathrm{lim}=19.5$ and Galactic-extinction-corrected $r=22.5-2.5\log_{10}(\mathrm{FLUX\_R})-2.165E(B-V)$ from the VAC. Within each population and redshift bin, we selected the faintest 20\% (those above the 80th percentile in $r$) and took the linearly interpolated 95th percentile of $\log M_\mathrm{lim}$. This calibration used the broader reference sample and remained fixed after the analysis mass cut. The limits rise with redshift and are consistently higher for quiescent than for star-forming galaxies (Fig.~\ref{fig:completeness}).

The selected sample requires $\log M_*\geq\log M_{\mathrm{lim,Q}}$; we used one unrounded quiescent limit throughout each redshift bin, although completeness can vary within it. We applied this cut before calculating statistics or the counts used in fractions, including in the broader redshift bins used for the AGN analysis. When the limit fell within a mass bin, the results calculated from galaxies above it might differ from those plotted for the whole bin. The limit derived for quiescent galaxies is higher than the combined estimate in every redshift bin. At the lowest redshift, it excludes the two lowest stellar mass bins and retains only part of the next bin, with the retained range moving to higher masses at higher redshift.

The quiescent limit was derived from galaxies with positive SFRs and a match in the group catalogue. The line styles, letters, and opacity distinguish mass intervals wholly above, crossing, or wholly below the adopted limits (Fig.~\ref{fig:distribution}). The opacity reinforces these three categories and does not indicate a measured completeness percentage. The remaining consequences of the flux selection are consolidated in Sect.~\ref{sec:limitations}.

Halo mass threshold.
We note that the classification of the $\log M_h = 12.5$--$13.0$ range is debated in the literature.
This mass range represents a transitional zone between isolated field galaxies and virialised groups \citep{Yang2007}.
Physically, haloes above $M_h \sim 10^{12}\,\mathrm{M}_\odot$ develop stable virial shocks that heat accreting gas to the virial temperature \citep{DekelBirnboim2006, Keres2005}, marking the onset of hot-mode accretion that is characteristic of group environments.
The Yang et al.\ group finder achieved a reliable group identification (${>}60\%$ member completeness in ${\sim}90\%$ of haloes) only above $\log M_h \gtrsim 12.5$ \citep{Yang2021}, which motivated our lower mass threshold.
Observationally, this regime can be classified as either ``field'' when using isolation criteria (richness $N=1$) or ``poor groups'' when satellite populations are present \citep{Weinmann2006, Wetzel2012}.
We adopted ``field'' for the $\log M_h < 13$ regime to establish a baseline for the comparison with cluster environments.

The mock catalogue tests showed that the Yang group finder recovers halo masses that are unbiased with respect to true halo masses with uncertainties of $\sim$0.2~dex at the high-mass end and $\sim$0.45~dex at the low-mass end \citep{Yang2021}.
For our DESI parent sample, we verified that velocity dispersions of group members scaled with halo mass as expected: the median $\sigma_v$ increases from $\sim$107~km\,s$^{-1}$ at $\log M_h = 12.5$--$13$ to $\sim$483~km\,s$^{-1}$ at $\log M_h \geq 14.5$ (groups with $\geq$5 sample members), consistent with the expected scaling for virialised systems.
This dynamical check supports the halo-mass ordering.

\subsubsection{Analysis bins and sample distribution}
\begin{itemize}
\item We divided the sample into five redshift bins: $z=0.05$--$0.15$, $0.15$--$0.25$, $0.25$--$0.35$, $0.35$--$0.45$, and $0.45$--$0.55$.
\item We used five stellar mass bins: $\log(M_*/\mathrm{M}_\odot) = 9.0$--$9.5$, $9.5$--$10.0$, $10.0$--$10.5$, $10.5$--$11.0$, and $11.0$--$12.0$.
\end{itemize}

\begin{figure*}
\centering
\includegraphics[width=0.95\textwidth]{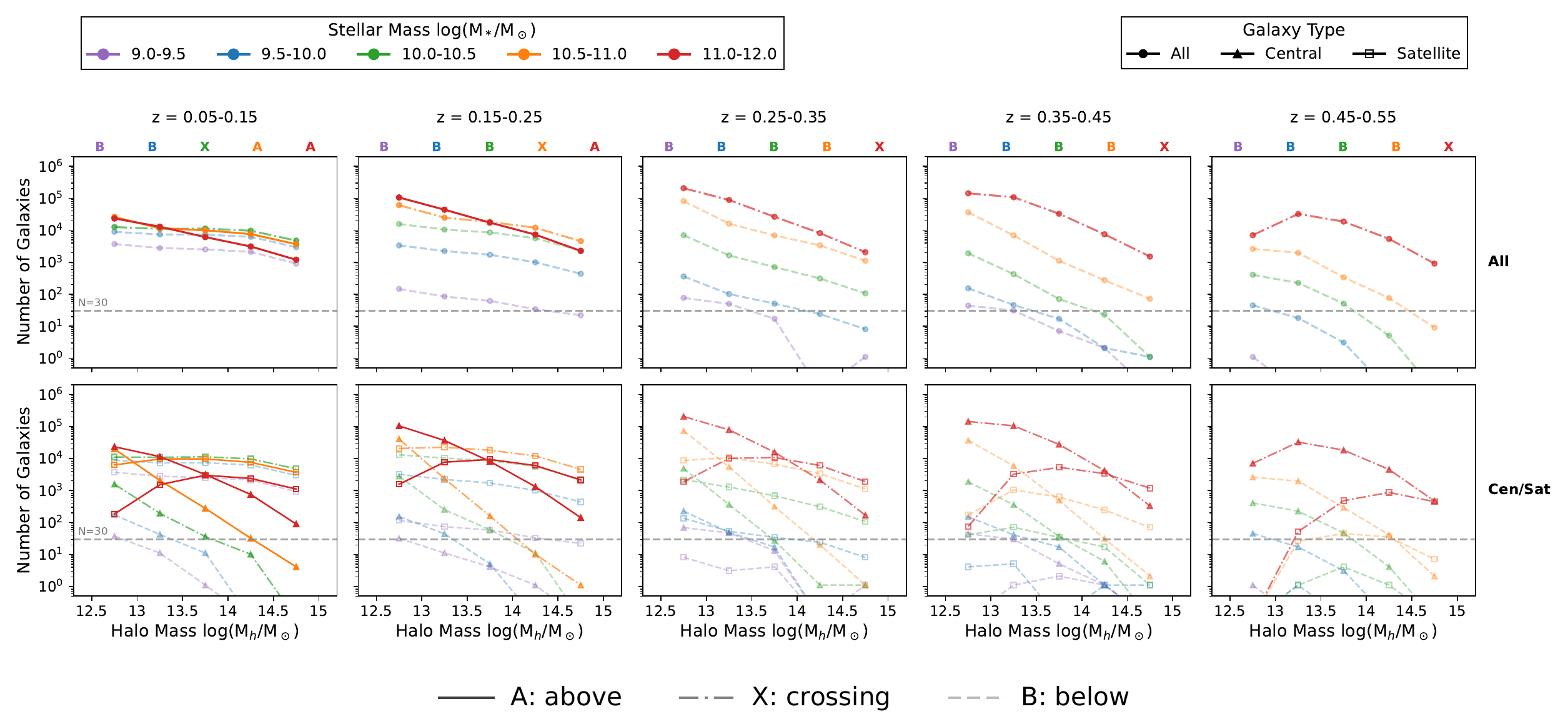}
\caption{Halo mass distribution of the observed parent sample in five redshift bins (columns). The rows show all-galaxy counts and central (triangles) and satellite (open squares) counts. The colours denote the stellar mass bins. Solid lines with A labels, dash-dotted lines with X labels, and dashed lines with B labels mark intervals wholly above, crossing, or wholly below the adopted completeness boundary, using both mass edges and the quiescent threshold of each redshift bin. The letters and progressively reduced opacity reinforce these categories. The positive counts remain visible even when they are below $N=30$.}
\label{fig:distribution}
\end{figure*}

Figure~\ref{fig:distribution} shows the observed parent halo mass distribution.
Both samples have their largest share in the field and a smaller cluster tail.

\section{Methods} \label{sec:methods}

\subsection{Specific star formation rate}

The specific star formation rate is calculated as
\begin{equation}
\mathrm{sSFR} = \frac{\mathrm{SFR\_CG}}{\mathrm{MASS\_CG}}\quad [\mathrm{yr}^{-1}],
\end{equation}
where \texttt{SFR\_CG} is in $\mathrm{M}_\odot\,\mathrm{yr}^{-1}$ and \texttt{MASS\_CG} in $\mathrm{M}_\odot$.
We used the VAC SED-based values for the primary measurements.

\begin{figure*}
\centering
\includegraphics[width=\textwidth]{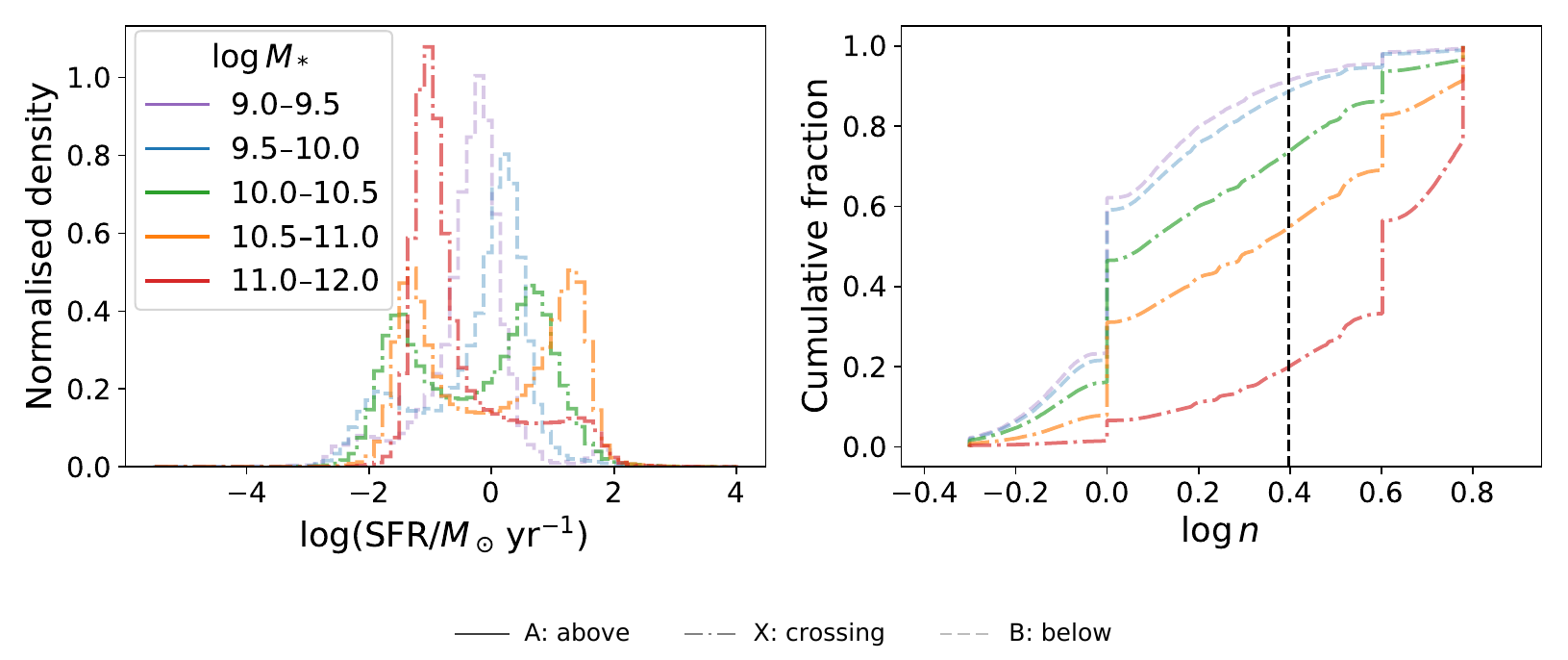}
\caption{Parent properties in five stellar mass bins. (a) Normalised $\log(\mathrm{SFR})$ distributions. (b) Cumulative $\log(n)$ distributions, with $n=2.5$ marked. Over the full redshift range, the $9$--$10$ bins lie below and higher-mass bins cross the completeness boundary. The opacity follows Fig.~\ref{fig:distribution}.}
\label{fig:sfr_distribution}
\end{figure*}

\begin{figure*}[!t]
\centering
\includegraphics[width=0.95\textwidth]{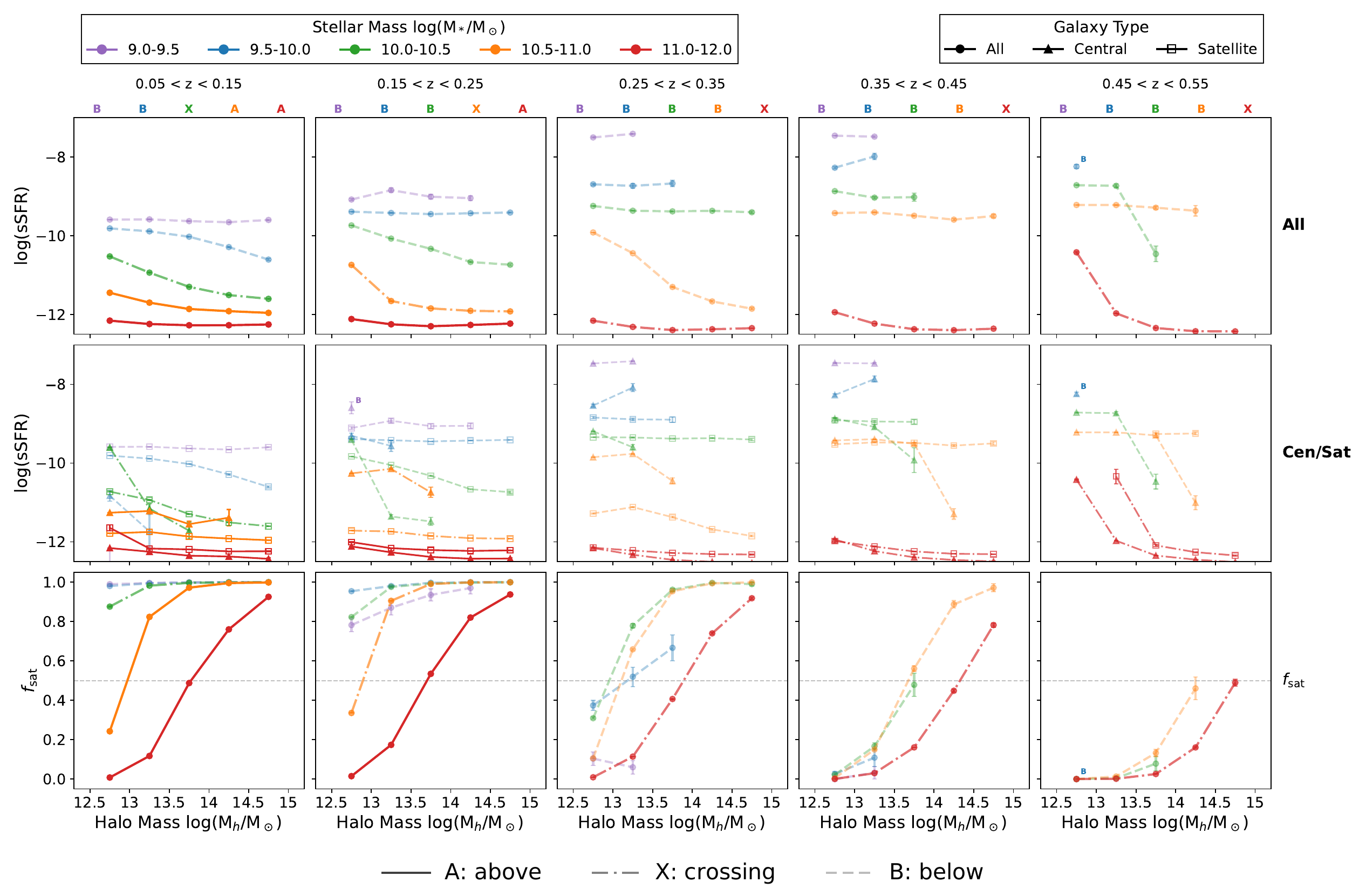}
\caption{Parent sSFR and satellite fraction vs. halo mass in five redshift bins (columns). The rows show the median sSFR for all galaxies, the median sSFR separately for centrals (triangles) and satellites (open squares), and the satellite fraction. The colours denote the stellar mass bins. The line styles and opacity follow Fig.~\ref{fig:distribution}. Each metric and population requires $N\geq30$. Bins with too few measurements are left blank.}
\label{fig:ssfr}
\end{figure*}

\begin{figure*}[!t]
\centering
\includegraphics[width=0.95\textwidth]{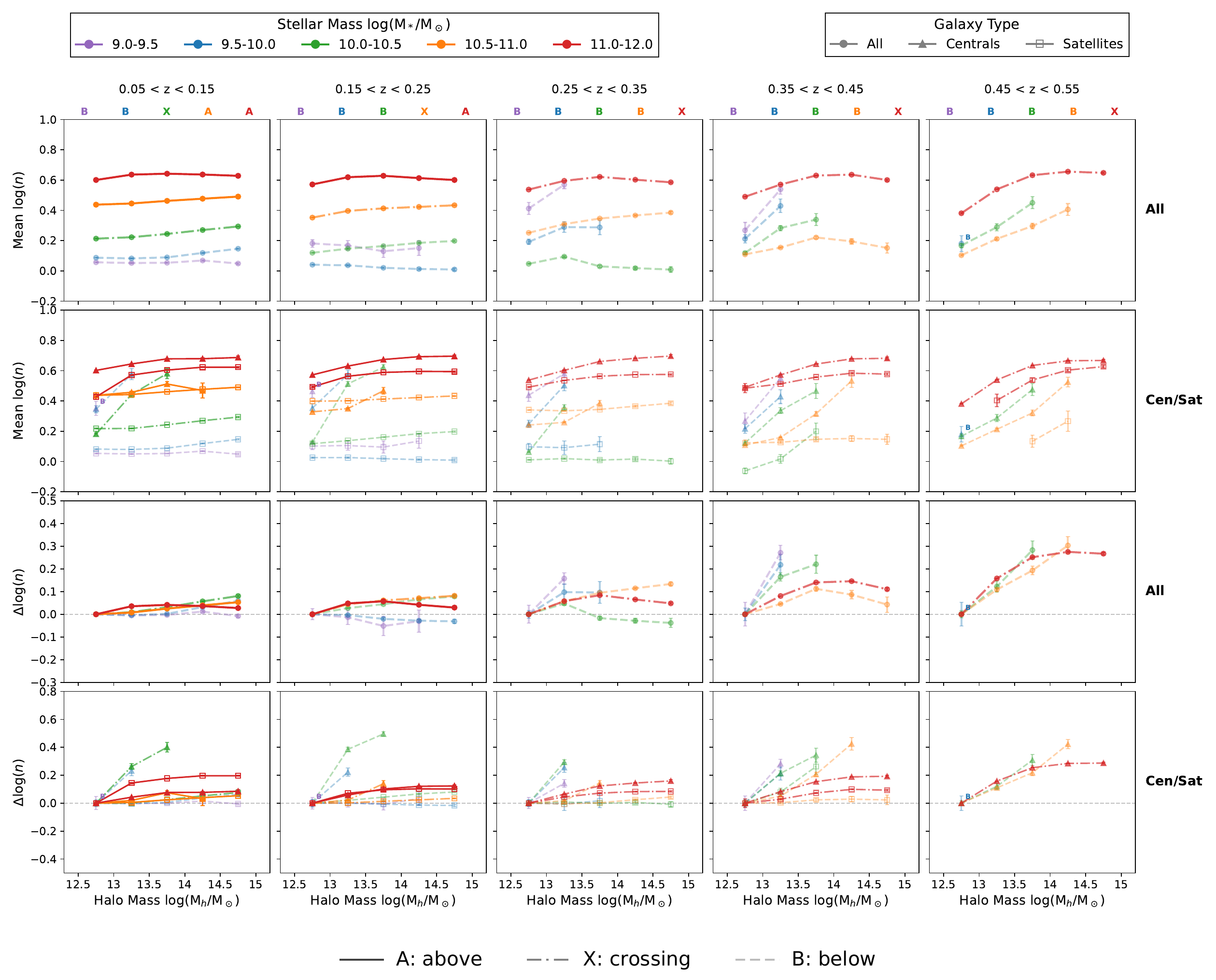}
\caption{Parent mean $\log(n)$ vs. halo mass in five redshift bins. Rows~1--2 show all galaxies and central (triangles) and satellite (open squares) populations. Rows~3--4 show their differences from the field. The line styles and opacity follow Fig.~\ref{fig:distribution}. Each mean requires at least 30 valid $0<n<8$ measurements. The differences also require a field mean meeting this threshold. The difference error bars omit the uncertainty in the field reference (see the methods section).}
\label{fig:sersic}
\end{figure*}

\begin{figure*}[!t]
\centering
\includegraphics[width=0.95\textwidth]{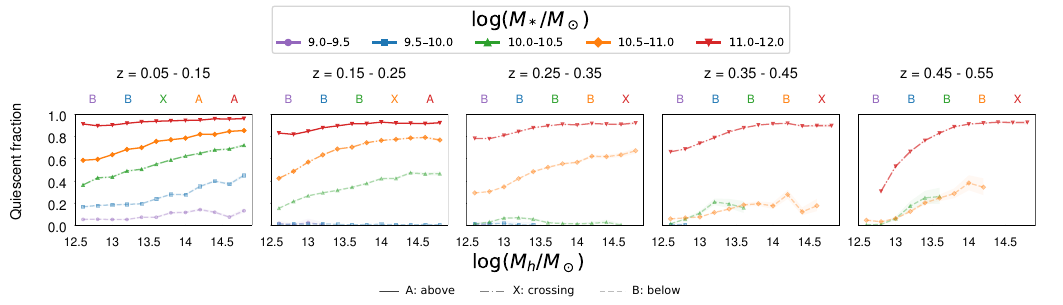}
\caption{Parent quiescent fraction $f_\mathrm{Q}$ ($\log\mathrm{sSFR}<-11$) vs. halo mass in five redshift bins, arranged from left to right. The colours and markers denote the stellar mass bins. The error bars show the Wilson $1\sigma$ intervals. Each point requires $N\geq30$. The line styles and opacity follow Fig.~\ref{fig:distribution}. The comparisons in the text use the selected sample.}
\label{fig:fQ}
\end{figure*}

\begin{figure*}[!t]
\centering
\includegraphics[width=0.95\textwidth]{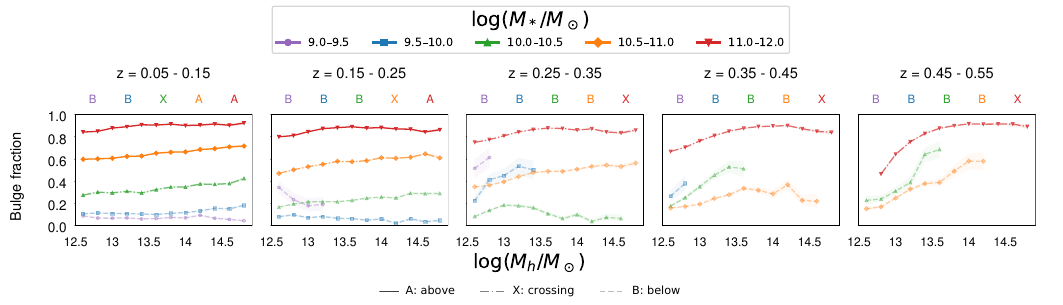}
\caption{Parent bulge-dominated fraction ($n>2.5$) vs. halo mass, with panels, mass colours, markers, and Wilson intervals as in Fig.~\ref{fig:fQ}. The line styles and opacity follow Fig.~\ref{fig:distribution}. The fractions require at least 30 valid S\'ersic measurements ($n>0$).}
\label{fig:fbulge}
\end{figure*}

\begin{figure*}[!t]
\centering
\includegraphics[width=0.95\textwidth]{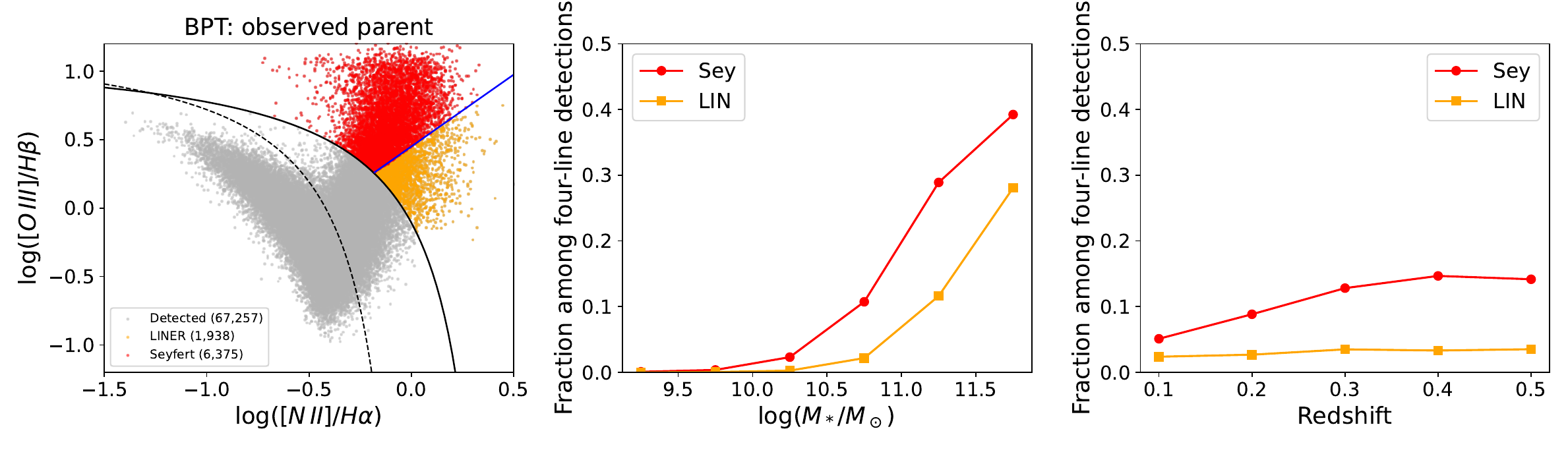}
\caption{BPT classification of parent galaxies with S/N$>3$ in all four lines. Panel~1: Line-ratio diagram with Seyferts (red), LINERs (orange), and Kewley (solid black), \citet{Kauffmann2003} (dashed black), and Schawinski (blue) classification boundaries. Panels~2--3: Seyfert and LINER fractions vs. stellar mass and redshift, requiring at least 30 four-line detections per bin. The mass panel combines redshifts, and the redshift panel combines stellar masses. Both panels include masses below the completeness limit and only galaxies with detected lines. The parent and detected counts appear in Fig.~\ref{fig:agndenom}.}
\label{fig:bpt}
\end{figure*}

\begin{figure*}[!t]
\centering
\includegraphics[width=0.95\textwidth]{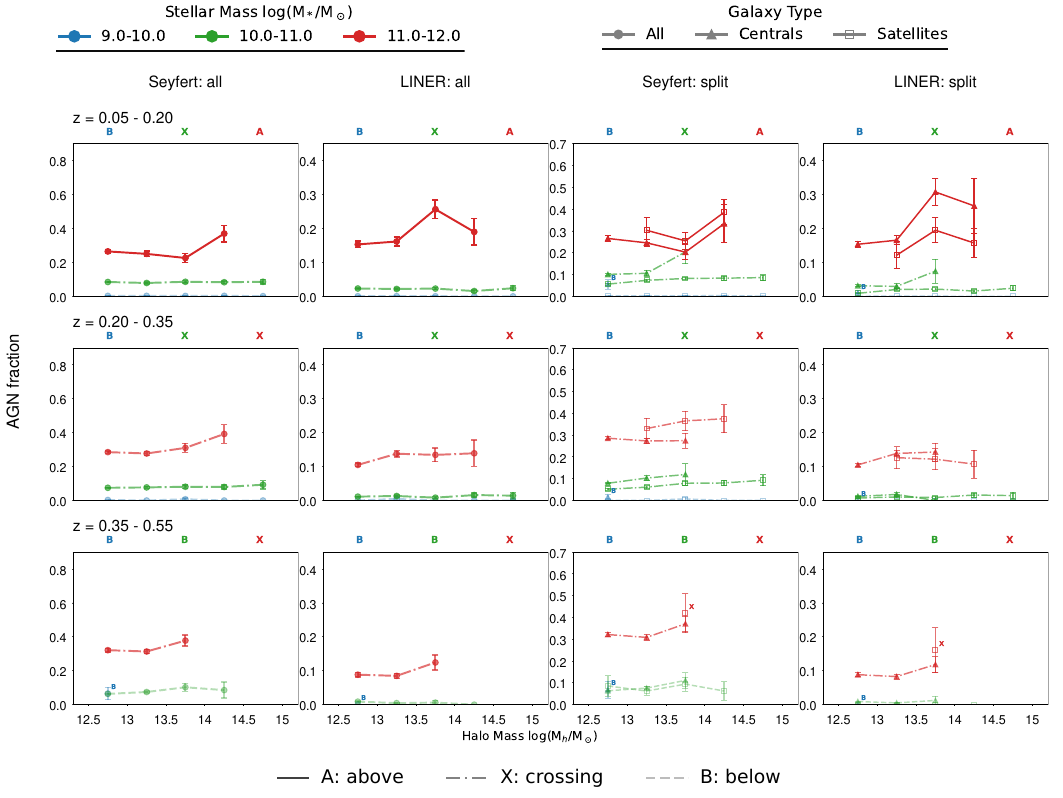}
\caption{Parent AGN fractions vs. halo mass in three redshift intervals: $0.05$--$0.20$, $0.20$--$0.35$, and $0.35$--$0.55$. The three rows show the redshift intervals in ascending order. The four columns show Seyfert fractions for all galaxies, LINER fractions for all galaxies, Seyfert fractions split into centrals (triangles) and satellites (open squares), and the corresponding LINER split. The colours denote $\log M_*$ bins $9$--$10$, $10$--$11$, and $11$--$12$. The line styles and opacity follow Fig.~\ref{fig:distribution}, using every original-bin threshold within each combined interval. Each population requires at least 30 four-line detections. All fractions use galaxies with detected lines.}
\label{fig:agn}
\end{figure*}

\begin{figure*}[!t]
\sidecaption
\includegraphics[width=0.65\textwidth]{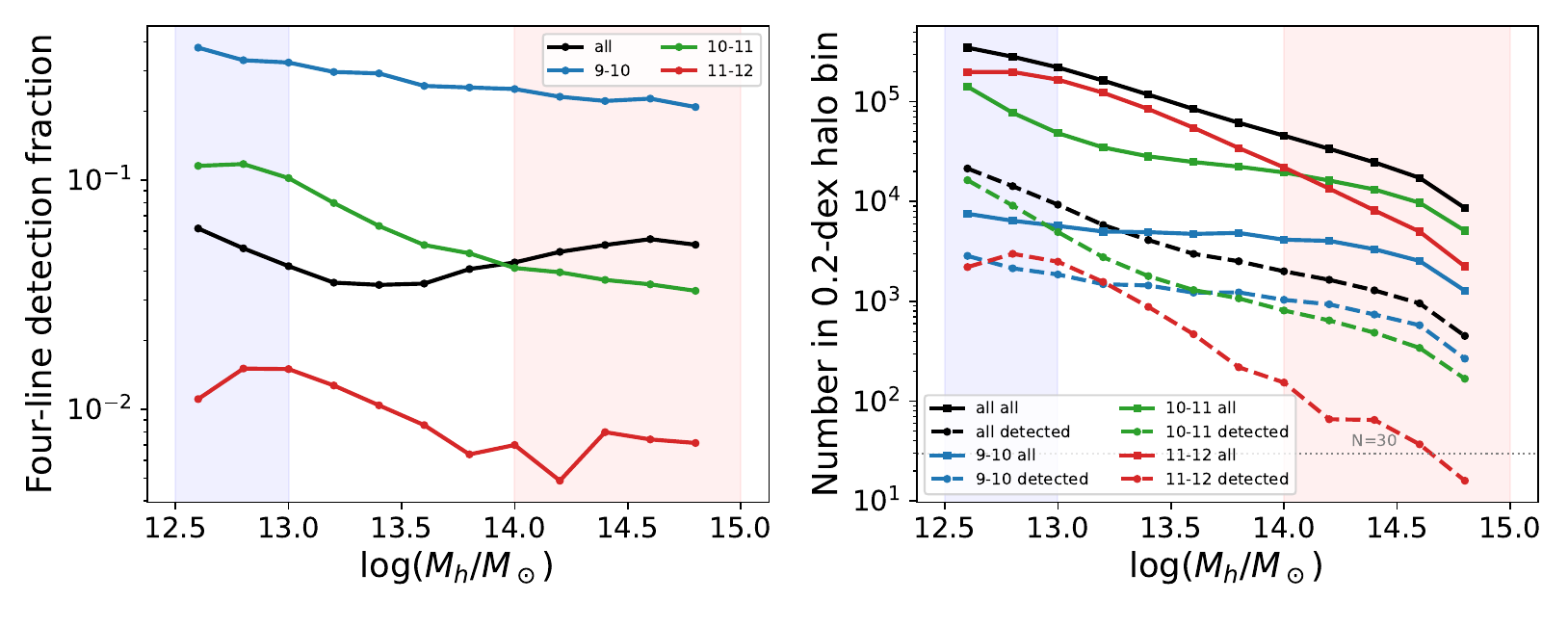}
\caption{Parent galaxies meeting the emission-line requirements. (a) Four-line detection fraction vs. halo mass, requiring at least 30 parent galaxies per bin. (b) Parent (solid) and four-line-detected (dashed) counts, retaining positive counts below $N=30$. These styles distinguish parent counts from four-line-detected counts, not completeness. Over the pooled redshift range, mass $9$--$10$ lies below the completeness boundary, while $10$--$11$ and $11$--$12$ cross it. The curve combining all stellar mass bins mixes these regimes. The four-line-detection requirement is defined in Sect.~\ref{sec:methods}.}
\label{fig:agndenom}
\end{figure*}

\subsection{SFR estimator choice and robustness}

Our primary SFR estimator was the VAC SED-based value (\texttt{SFR\_CG}) to maintain internal consistency across the full sample.
As a check, we computed the SFRs from the catalogue Balmer fluxes using \citet{Kennicutt1998} and the dust prescriptions in Appendix~\ref{app:balmer}, retaining the VAC continuum-subtraction treatment of stellar absorption \citep{ZouHu2024}.
Full details of the Balmer comparison, including the sample characterisation, method-split offsets, and figure diagnostics, are given in Appendix~\ref{app:balmer}.

\subsection{AGN classification}

We classified the AGNs using the [N~{\sc ii}] Baldwin--Phillips--Terlevich (BPT) diagram \citep{Baldwin1981} with emission line fluxes from the stellar mass VAC.
We required a signal-to-noise ratio (S/N) $>3$ in all four BPT lines (H$\alpha$, H$\beta$, [N~{\sc ii}]~$\lambda6583$, and [O~{\sc iii}]~$\lambda5007$) for the AGN classification. This yielded 67,257 parent and 16,091 selected four-line detections. We made no further broad-line rejection after the quasar exclusion.

The \citet{Kewley2001} maximum starburst line separates AGN from star-forming galaxies,
\begin{equation}
\log([\mathrm{O\,III}]/\mathrm{H}\beta) = \frac{0.61}{\log([\mathrm{N\,II}]/\mathrm{H}\alpha) - 0.47} + 1.19.
\end{equation}
Galaxies above this line are classified as AGN.
The \citet{Schawinski2007} line separates Seyferts from low-ionisation nuclear emission-line regions (LINERs),
\begin{equation}
\log([\mathrm{O\,III}]/\mathrm{H}\beta) = 1.05 \times \log([\mathrm{N\,II}]/\mathrm{H}\alpha) + 0.45,
\end{equation}
with Seyferts above and LINERs below this line.
The AGN fractions were calculated as $f_\mathrm{AGN} = N_\mathrm{AGN} / N_\mathrm{em-line}$, where the denominator is galaxies with S/N~$>3$ in all four BPT lines.

\subsection{Morphological analysis}

The galaxy morphology was characterised using the logarithm of the S\'{e}rsic index, $\log(n)$, where $n \sim 1$ ($\log(n) \approx 0$) indicates disc-dominated systems, and $n \sim 4$ ($\log(n) \approx 0.6$) indicates elliptical galaxies.
The mean $\log(n)$ requires $0<n<8$; bulge fractions require a positive, finite S\'ersic index. Using $\log(n)$ gives equal weight to equal multiplicative intervals across the broad index distribution.
The S\'{e}rsic index difference relative to the field is
\begin{equation}
\Delta \log(n) = \langle \log(n) \rangle_{M_h} - \langle \log(n) \rangle_\mathrm{field}.
\end{equation}

\subsection{Statistical threshold}

The plotted measurements required at least 30 parent-sample galaxies for each population and measured quantity. For measurements that were restricted to galaxies above the completeness limit, this requirement applied after the cut. Field--cluster contrasts require the count threshold in both environments. The panels showing galaxy counts retain positive values below the threshold, and curves leave gaps in bins with too few measurements.

The uncertainties in the quiescent and bulge-dominated fractions were estimated using Wilson binomial intervals \citep{Brown2001}; other fraction errors use galaxy-level binomial estimates. The median-sSFR error bars (Fig.~\ref{fig:ssfr}) use $1.4826\,\mathrm{MAD}/\sqrt{N}$, where MAD is the median absolute deviation. The mean $\log(n)$ error bars (Fig.~\ref{fig:sersic}) use $\mathrm{SD}/\sqrt{N}$, where SD is the sample standard deviation; the difference bars in that figure omit the uncertainty in the field reference.
The field--cluster sSFR and $\log(n)$ errors for both samples, including Fig.~\ref{fig:phase}, use standard deviations of 500 bootstrap resamples of individual galaxies.

\section{Results} \label{sec:results}

The figures show measurements for the parent sample, including bins affected by incompleteness. Our main inferences about the galaxy population are based on comparisons above the completeness limit, with the measurement requirements given in Sect.~\ref{sec:methods} and the interpretive limits consolidated in Sect.~\ref{sec:limitations}.

\subsection{Star formation versus environment}

Figure~\ref{fig:ssfr} shows the parent median sSFR.
The sSFR trends with halo mass partly reflect the growing proportion of quenched galaxies.

In the selected sample, the environmental dependence at low-to-intermediate redshifts ($z < 0.25$) is strongest at intermediate stellar masses ($\log(M_*/\mathrm{M}_\odot) \sim 10$--$11$), where the median sSFR can transition from star-forming values in the field to quiescent levels in massive clusters.
In the highest stellar mass bin ($\log(M_*/\mathrm{M}_\odot) = 11.0$--$12.0$), galaxies are predominantly quenched in field and cluster environments, leaving little environmental contrast. At $0.25\leq z<0.45$, only this bin contains enough selected galaxies for field--cluster comparisons, and the contrasts in both redshift intervals are small. In the selected sample, the median sSFR is lower in clusters in every field--cluster comparison meeting the count requirement.
At the lowest stellar masses ($\log(M_*/\mathrm{M}_\odot)<10$), no galaxies remain after the adopted completeness cut was applied.

When the selected sample is split into centrals and satellites, satellites are more quenched than centrals at fixed stellar mass and halo mass in the well-populated intermediate-mass regimes, but not uniformly in every mass--redshift bin.
At intermediate masses ($\log(M_*/\mathrm{M}_\odot)=10.5$--$11.0$) and $z=0.15$--$0.25$, the median satellite sSFR sits at quiescent levels while centrals remain star forming, consistent with environmental quenching acting primarily on the satellite population.
At the highest stellar masses ($\log(M_*/\mathrm{M}_\odot)>11$), the two populations are broadly quiescent, although field satellites have a higher median sSFR than field centrals in some retained redshift bins.

At intermediate masses ($\log(M_*/\mathrm{M}_\odot)=10.5$--$11.0$) and $z<0.25$, the selected satellite fraction increases toward clusters. This changing mix contributes to the combined sSFR trend without requiring stronger suppression within each population.

\subsection{S\'{e}rsic index versus environment}

The parent S\'{e}rsic index distributions (Fig.~\ref{fig:sfr_distribution}b) show the observed morphology--mass relation across the stellar mass bins.
Figure~\ref{fig:sersic} shows the parent mean $\log(n)$ as a function of halo mass. Above the completeness limit, galaxies in clusters have slightly higher S\'{e}rsic indices than field galaxies at fixed stellar mass and redshift.

\subsection{Quiescent and bulge-dominated fractions versus environment}
\label{sec:fractions}

To assess whether environmental changes are clearer in population fractions than in the median sSFR or mean logarithmic S\'ersic index, we examined the quiescent and bulge-dominated fractions \citep[following the median-versus-fraction diagnostic of][]{Peng2010}. We used the same halo mass, stellar mass, and redshift bins as in the preceding sections. We defined quiescent galaxies as
$\log(\mathrm{sSFR}/\mathrm{yr}^{-1}) < -11$ and bulge-dominated galaxies as S\'ersic index $n > 2.5$, the
same boundary as adopted in Fig.~\ref{fig:sfr_distribution}b.

Figure~\ref{fig:fQ} shows the parent quiescent fractions. In the selected sample, $f_\mathrm{Q}$ is higher in clusters than in the field in every stellar mass and redshift bin used for comparison. In the first two redshift bins ($0.05\leq z<0.25$), the field--cluster difference in $f_\mathrm{Q}$ is smaller in the highest stellar mass bin than in the intermediate-mass bins. Most galaxies in the highest-mass bin are already quiescent in both environments.

At intermediate masses, the median sSFR shifts sharply when quiescent galaxies become the majority, without requiring exceptionally strong suppression of each galaxy. The smoother $f_\mathrm{Q}$ trend, together with the changing satellite fraction, makes the effect of the changing population mix explicit.

Figure~\ref{fig:fbulge} shows parent bulge fractions. This statistic is often considered more environment-sensitive than mean S\'ersic index \citep{vanderWel2008, Bamford2009}. In every selected field--cluster comparison meeting the count requirement, $f_\mathrm{bulge}$ is higher in clusters, but its increase is smaller than that of $f_\mathrm{Q}$ at intermediate masses.

\subsection{AGN fraction versus environment}

Figure~\ref{fig:bpt} shows the BPT classification of parent galaxies. Its middle panel combines redshift bins to show the Seyfert and LINER fractions versus stellar mass. Both fractions rise with stellar mass, and together, they exceed one half in the highest mass bin.

Figure~\ref{fig:agn} shows the AGN fractions in the parent sample in three broader redshift intervals. Seyferts are generally more common than LINERs. At intermediate masses ($\log(M_*/\mathrm{M}_\odot)=10$--$11$), the separate redshift panels show no clear monotonic trend with halo mass for either class.

At the highest masses ($\log(M_*/\mathrm{M}_\odot)=11$--$12$), the Seyfert curves rise in some massive-halo bins at low to intermediate redshift, but these bins are sparse and uncertain. The LINER fractions vary without a consistent progression with halo mass.

Figure~\ref{fig:agndenom} shows how many parent galaxies meet the four-line detection requirement. In the displayed parent sample, the detection fraction most clearly declines towards clusters at intermediate masses.

\begin{figure*}[!t]
\centering
\includegraphics[width=0.98\textwidth]{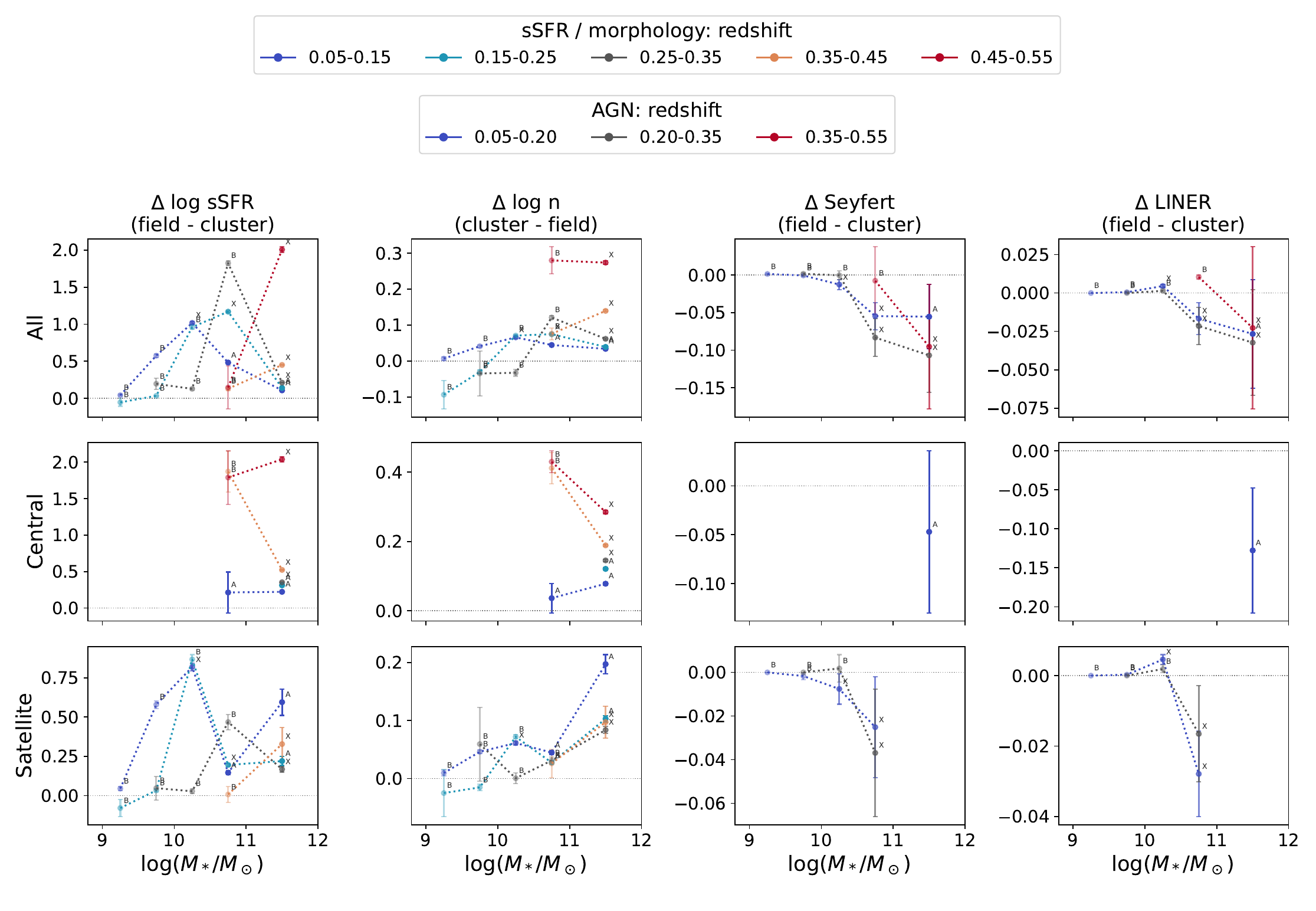}
\caption{Parent field--cluster differences vs. stellar mass. The rows show all galaxies, centrals, and satellites. Columns: $\Delta$sSFR (field minus cluster), $\Delta\log(n)$ (cluster minus field), and $\Delta$Seyfert and $\Delta$LINER (field minus cluster). The colours denote redshift; the field is $12.5\leq\log M_h<13$ and the cluster $14\leq\log M_h<16$. Each metric requires at least 30 objects in each environment. The A, X, and B point letters and marker and error-bar opacity follow Fig.~\ref{fig:distribution}. The dotted connectors are visual guides only. The sSFR and morphology bars show bootstrap standard deviations. The AGN bars show the propagated binomial errors (Sect.~\ref{sec:methods}).}
\label{fig:phase}
\end{figure*}

\subsection{Field versus cluster across galaxy properties}

Figure~\ref{fig:phase} summarises the parent contrasts between field ($\log M_h<13$) and cluster ($\log M_h\geq14$) samples at fixed stellar mass.
Specifically, we computed the
$\Delta\mathrm{sSFR} = \mathrm{median}(\log\,\mathrm{sSFR})_\mathrm{field} - \mathrm{median}(\log\,\mathrm{sSFR})_\mathrm{cluster}$ (positive values indicate suppressed star formation in clusters);
$\Delta\log(n) = \mathrm{mean}(\log\,n)_\mathrm{cluster} - \mathrm{mean}(\log\,n)_\mathrm{field}$ (positive values indicate higher S\'{e}rsic indices in clusters);
and $\Delta f_\mathrm{AGN} = f_\mathrm{AGN,field} - f_\mathrm{AGN,cluster}$ for Seyfert and LINER fractions (positive values indicate higher AGN fractions in the field).

The sSFR contrasts in Fig.~\ref{fig:phase} are mostly positive: at fixed stellar mass and redshift, the median sSFR is generally lower in clusters than in the field. For centrals and satellites combined, intermediate-mass galaxies show a larger field--cluster sSFR difference than galaxies in the highest-mass bin within each of the first two redshift intervals ($0.05$--$0.15$ and $0.15$--$0.25$). The positive contrasts in the separate central and satellite panels show that the environmental difference is also present within each population.

The morphological contrasts are broadly comparable between populations at low to intermediate redshift.
In the parent sample we show, the Seyfert contrasts tend to be negative at intermediate and high masses, corresponding to higher fractions in clusters, whereas LINER contrasts vary in sign except at the highest masses, where they are also negative in every comparison, but within $2\sigma$ of zero. In the selected sample, the all-galaxy Seyfert contrast is likewise negative in all five comparisons meeting the count requirement.

\section{Discussion} \label{sec:discussion}

\subsection{Main trends and the central and satellite mix}

Three main trends emerge from comparisons above the completeness limit (Sect.~\ref{sec:results}).
First, the median sSFR decreases with increasing halo mass, with a stronger field--cluster contrast in specific stellar mass and redshift regimes.
Second, the mean $\log(n)$ increases towards higher halo mass with a modest field-to-cluster contrast.
Third, among galaxies with all four diagnostic lines detected, Seyfert and LINER fractions show no strong environmental dependence; the Seyfert contrasts tend to be negative at intermediate and high masses, and both classes are marginally higher in clusters at the highest masses. The fixed-mass comparisons reduce the bias from the differences in the masses of line-detected galaxies.

At intermediate stellar mass and $z<0.25$, the satellite fraction increases with halo mass, and satellites have a lower sSFR than centrals at fixed stellar mass and halo mass.
Therefore, part of the apparent environmental suppression in the combined sample reflects the changing central and satellite mixture, not only additional in situ suppression within each population.
The quiescent- and bulge-fraction results reinforce this interpretation (Sect.~\ref{sec:fractions}): the quiescent fraction responds most strongly at intermediate masses above the completeness limit, consistent with \citet{Peng2010}. The weaker morphological response is consistent with but does not establish quenching before structural transformation.

\subsection{Physical interpretation}

These results are consistent with environmental processes and changes in the mix of galaxy populations.
The lower sSFR of satellites relative to centrals at intermediate stellar masses is consistent with mechanisms such as ram-pressure stripping and strangulation in dense haloes (e.g. \citealt{GunnGott1972, Larson1980}; see also \citealt{Cakir2025} for resolved quenching signatures).
The increase in $\log(n)$ with halo mass suggests that structural transformation and quenching are linked at least statistically, although our data do not by themselves establish causal ordering.
At the highest stellar masses, weak additional environmental contrast is plausibly related to the dominance of already-quenched populations, so differential environmental signatures become smaller in relative terms.

\subsection{Comparison of SFR estimators}
\label{sec:sfrcompare}

The Balmer and SED SFRs have different zero points, with offsets that differ between the two Balmer methods (Appendix~\ref{app:balmer}). For galaxies above the completeness limit with both estimates, the field--cluster differences have the same direction in every comparison meeting the count requirement, although their amplitudes differ. Selecting galaxies with Balmer measurements generally weakens the apparent environmental suppression and changes the central and satellite mix. Figure~\ref{fig:appendix_balmer_bridge} separates this sample-selection effect from the estimator difference.

\subsection{Limitations}
\label{sec:limitations}

The BGS\_BRIGHT flux limit remains the main sample limitation. The adopted quiescent-galaxy threshold reduces colour-dependent incompleteness but does not guarantee 95\% recovery of every quiescent population. It is calibrated from galaxies with positive SFRs and group-catalogue matches, and completeness can vary within a redshift bin. The count requirement does not correct incompleteness. Preferentially excluding quiescent galaxies can raise the measured sSFR and lower the quiescent fraction, while field--cluster differences can be biased in either direction in below-limit and crossing bins. Because the completeness cut removes the low-mass comparisons, these data cannot establish whether the characteristic mass of environmental quenching evolves with redshift \citep{Peng2010, Wetzel2012}. Volume-limited subsamples would provide a direct check of the remaining selection effects.

Group-based quantities add further uncertainty. Abundance matching reduces the redshift dependence of raw richness, but halo mass scatter \citep{Yang2021} and projected interlopers remain. The adopted field interval also includes transitional low-mass groups rather than the most isolated galaxies. We did not independently measure errors in central and satellite labels or group membership, so their effect on population-specific contrasts is unquantified. The bootstrap errors resample individual galaxies and do not include correlations among members of the same group.

The magnitude of the sSFR differences depends on the estimator and available galaxy subset. Very low SED SFRs in quiescent galaxies are model dependent rather than detections of ongoing star formation, while Balmer measurements preferentially select line-emitting galaxies. Their comparison is additionally affected by dust corrections and the assumption that line emission follows the $r$-band continuum in the aperture correction. The H$\beta$ proxy additionally uses stellar attenuation to estimate gas attenuation, and its small high-redshift sample is too sparse for general environmental comparisons. Agreement in direction does not establish an estimator-independent suppression amplitude.

The BPT fractions are conditional on the detection of all four diagnostic lines, whose frequency varies with environment and stellar mass. Pooling broad mass or redshift ranges can therefore change the composition of the detected sample. Their galaxy-level binomial errors understate the uncertainty where a fraction is close to zero, so the smallest LINER error bars in Fig.~\ref{fig:phase} should not be over-interpreted. The selection misses line-weak, obscured, and some broad-line activity, while the LINER classifications can include ionisation by evolved stars; no H$\alpha$ equivalent-width cut was applied \citep{CidFernandes2011}. The resulting fractions do not measure total AGN incidence. They also average over whole haloes and cannot test cluster-core behaviour. Infrared, X-ray, or radio AGN selections would address the detection limits, while positions and velocities relative to group centres would test the core-dependent trends. At high redshift and high mass, the selected satellite and AGN subsamples are too sparse for reliable population-split comparisons.

Finally, the analysis is differential and statistical. The observed associations do not establish the causal order of quenching and structural change or determine quenching timescales for individual galaxies.

\section{Conclusions} \label{sec:conclusions}

We have examined the environmental dependence of star formation, morphology, and AGN activity using about 1.4 million BGS\_BRIGHT galaxies from DESI DR1 at $0.05 < z < 0.55$.
The principal conclusions below are supported by comparisons above the adopted completeness limit; the parent sample was analysed across the observed mass range.

\begin{enumerate}
\item The selected comparisons show a lower median sSFR in clusters than in the field in all well-populated stellar mass and redshift bins above or crossing the completeness limit. At intermediate stellar masses and $z<0.25$, satellites show lower sSFR than centrals at fixed stellar mass and halo mass, and their rising fraction contributes to the combined environmental trend.

\item The S\'{e}rsic index increases only weakly from field to cluster at fixed stellar mass and redshift.

\item Among selected galaxies detected in all four BPT lines, Seyfert and LINER fractions show no strong environmental dependence at fixed stellar mass; both are at most marginally higher in clusters, with the largest but most uncertain differences at the highest stellar masses.
\end{enumerate}

\section*{Data availability}
Machine-readable tables accompanying this manuscript provide paired parent and selected measurements, the fixed completeness calibration, diagnostics for centrals and satellites, and Balmer diagnostics. The accompanying \texttt{ReadMe} documents their mapping, units, and measurement requirements. The machine-readable tables are available in Zenodo at \url{https://doi.org/10.5281/zenodo.23157951}. Public inputs are DESI DR1 spectra and redshifts, the stellar mass VAC, and the \citet{Yang2021} DR9 group catalogue.

\begin{acknowledgements}
We thank the anonymous referee for a careful and constructive report that improved the clarity and rigour of this paper.
D.K.\ acknowledges support from the National Research Foundation of Korea (NRF) grant funded by the Korean government (MSIT) (No.\ NRF-2022R1C1C2004506).

This research used data obtained with DESI.
DESI construction and operations are managed by the Lawrence Berkeley National Laboratory.
This material is based upon work supported by the U.S.\ Department of Energy, Office of Science, Office of High-Energy Physics, under Contract No.\ DE--AC02--05CH11231, and by the National Energy Research Scientific Computing Center, a DOE Office of Science User Facility under the same contract.
Additional support for DESI was provided by the U.S.\ National Science Foundation (NSF), Division of Astronomical Sciences under Contract No.\ AST-0950945 to the NSF's National Optical-Infrared Astronomy Research Laboratory;
the Science and Technology Facilities Council of the United Kingdom;
the Gordon and Betty Moore Foundation;
the Heising-Simons Foundation;
the French Alternative Energies and Atomic Energy Commission (CEA);
the National Council of Humanities, Science and Technology of Mexico (CONAHCYT);
the Ministry of Science and Innovation of Spain (MICINN),
and by the DESI Member Institutions: \url{https://www.desi.lbl.gov/collaborating-institutions}.
The DESI collaboration is honored to be permitted to conduct scientific research on I'oligam Du'ag (Kitt Peak), a mountain with particular significance to the Tohono O'odham Nation.
Any opinions, findings, and conclusions or recommendations expressed in this material are those of the author(s) and do not necessarily reflect the views of the U.S.\ National Science Foundation, the U.S.\ Department of Energy, or any of the listed funding agencies.

This work made use of the DESI DR1 stellar-mass and emission-line value-added catalogue \citep{ZouHu2024}.
We also made use of the extended halo-based group catalogue constructed for DESI DR1 \citep{Yang2021}.

This work made use of generative AI tools (Claude, Anthropic; GPT, OpenAI; Gemini, Google; Kimi, Moonshot AI) for assistance with manuscript drafting, data analysis code development, figure generation, and verification during revision. The author is responsible for the scientific content, interpretation, and conclusions.
\end{acknowledgements}

\bibliographystyle{aa}
\bibliography{kim2025_aa}

\begin{appendix}
\nolinenumbers

\section{Balmer-line robustness diagnostics}
\label{app:balmer}

This appendix presents the Balmer-line SFR comparison and two complementary diagnostics assessing the robustness of our SFR measurements with respect to estimator choice and sample selection.
The primary concern is whether the environmental trends reported in the main text are driven by the choice of SED-based versus Balmer-line SFR estimators, or by differences in the galaxy subsamples for which each method is applicable.

\subsection*{Balmer-line SFR prescription and offsets}

Following \citet{ZouHu2024}, we use H$\alpha$ luminosity where available and, at $z>0.43$ when H$\alpha$ is unusable, an H$\beta$ proxy. The H$\alpha$-based estimate uses non-negative Balmer-decrement reddening; the H$\beta$ proxy uses SED attenuation. We scale the Balmer luminosities to total $r$-band photometry using the catalogue aperture factor \citep{ZouHu2024}. We multiply the Salpeter-scale \citet{Kennicutt1998} SFRs by $f_{\rm IMF}=0.63$ to adopt a common Chabrier normalisation \citep{MadauDickinson2014}.
The H$\alpha$-based estimate requires H$\alpha$ and H$\beta$ at S/N$>5$. At $z>0.43$, where H$\alpha$ falls in the noisy red end of the spectrograph, we estimate the SFR from H$\beta$ when it is detected at S/N$>5$ and H$\alpha$ is too weak (S/N$\leq5$) or unavailable. The line-based SFR calibrations used are:
\begin{equation}
\mathrm{SFR}_{\mathrm{H}\alpha} = f_{\rm IMF}\,7.9\times10^{-42}\,L_{\mathrm{H}\alpha}\quad [\mathrm{M}_\odot\,\mathrm{yr}^{-1}],
\end{equation}
and, when H$\alpha$ is not usable,
\begin{equation}
\mathrm{SFR}_{\mathrm{H}\beta} = f_{\rm IMF}\,7.9\times10^{-42}\,\left(2.86\,L_{\mathrm{H}\beta}\right)\quad [\mathrm{M}_\odot\,\mathrm{yr}^{-1}],
\end{equation}
where the dust- and aperture-corrected Balmer luminosities $L_{\mathrm{H}\alpha}$ and $L_{\mathrm{H}\beta}$ are expressed in $\mathrm{erg\,s^{-1}}$, as required by the \citet{Kennicutt1998} calibration, and 2.86 is the Case-B intrinsic Balmer ratio used to convert H$\beta$ to an equivalent H$\alpha$ luminosity. Luminosity distances assume a flat $\Lambda$CDM cosmology with $H_0=70\,\mathrm{km\,s^{-1}\,Mpc^{-1}}$ and $\Omega_m=0.3$.
Dust attenuation for the H$\alpha$-based estimate uses the \citet{Cardelli1989} law with $k_{\mathrm{H}\alpha}=2.53$ and $k_{\mathrm{H}\beta}=3.61$, enforcing non-negative Balmer-decrement reddening ($E(B-V)\ge 0$). The H$\beta$ proxy retains the VAC SED-based $A_V$ with the \citet{Calzetti2000} law, $k_{\mathrm{H}\beta}=4.598$ and $R_V=4.05$. Stellar Balmer absorption is already corrected by the VAC continuum fit.

\begin{figure*}
\centering
\includegraphics[width=0.98\textwidth]{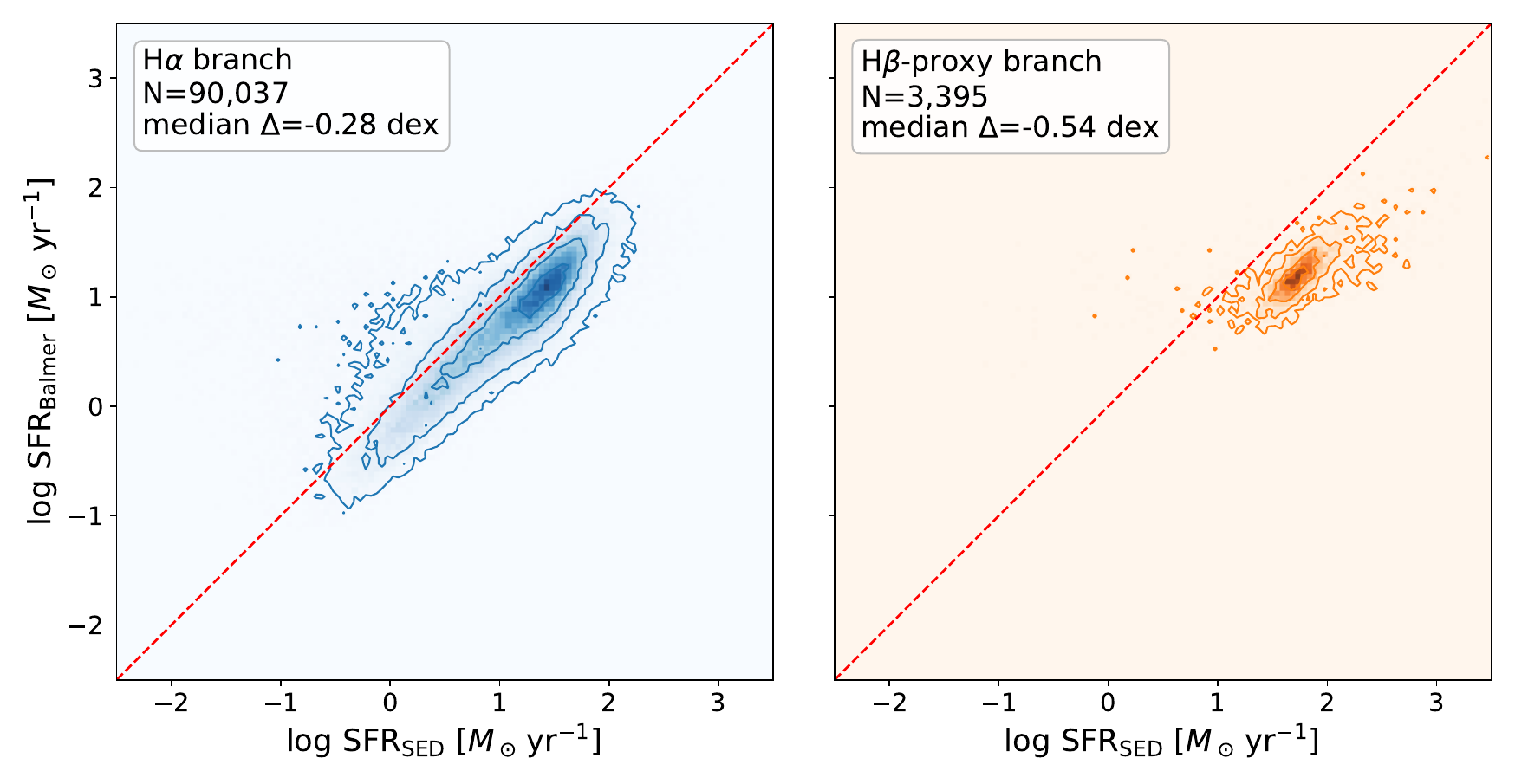}
\caption{SED versus aperture-corrected, Chabrier-normalised Balmer SFR for parent galaxies with both estimates: H$\alpha$ branch (left) and H$\beta$ proxy (right). Both panels use flux- and line-detected galaxies and include masses below the completeness limit; they compare estimators. Median offsets and environmental comparisons for galaxies above the completeness limit are discussed in the text.}
\label{fig:appendix_balmer_1to1_contour}
\end{figure*}

For galaxies above the adopted completeness limit with both SFR estimates at $z\leq0.49$, keeping the comparison within the redshift range where H$\alpha$ is covered by DESI, the median offset $\Delta\log\mathrm{SFR}\equiv\log\mathrm{SFR}_{\mathrm{Balmer}}-\log\mathrm{SFR}_{\mathrm{SED}}$ is approximately $-0.14$~dex for H$\alpha$ and $-0.45$~dex for the much smaller subset using the H$\beta$ proxy; the combined result is dominated by H$\alpha$.
For the broader sample in Fig.~\ref{fig:appendix_balmer_1to1_contour}, which also includes galaxies below the completeness limit, the median H$\alpha$ offset is $-0.28$~dex. This is similar in size and sign to the $-0.25$~dex reported for the higher-redshift Wide Field Camera 3 Infrared Spectroscopic Parallel survey sample by \citet{Henry2021}, whose comparison with the Cosmic Assembly Near-infrared Deep Extragalactic Legacy Survey shows little offset. The samples and modelling choices differ, so this resemblance does not establish a common cause. With SED estimates calibrated partly against emission-line SFRs, \citet{Salim2016} find close agreement for main-sequence galaxies without AGN.

\citet{Henry2021} associate the offset in their parallel-survey sample with higher SED dust estimates from limited photometric coverage. For nearby H\,\textsc{i}-deficient galaxies, \citet{Boselli2015} instead find SED SFRs below H$\alpha$, a discrepancy they argue is not explained by the adopted simple SED star-formation histories. These contrasting results show that the sign and cause of an estimator offset are not universal; the limitations of our estimator comparison are consolidated in Sect.~\ref{sec:limitations}.

\begin{figure*}
\centering
\includegraphics[width=0.98\textwidth]{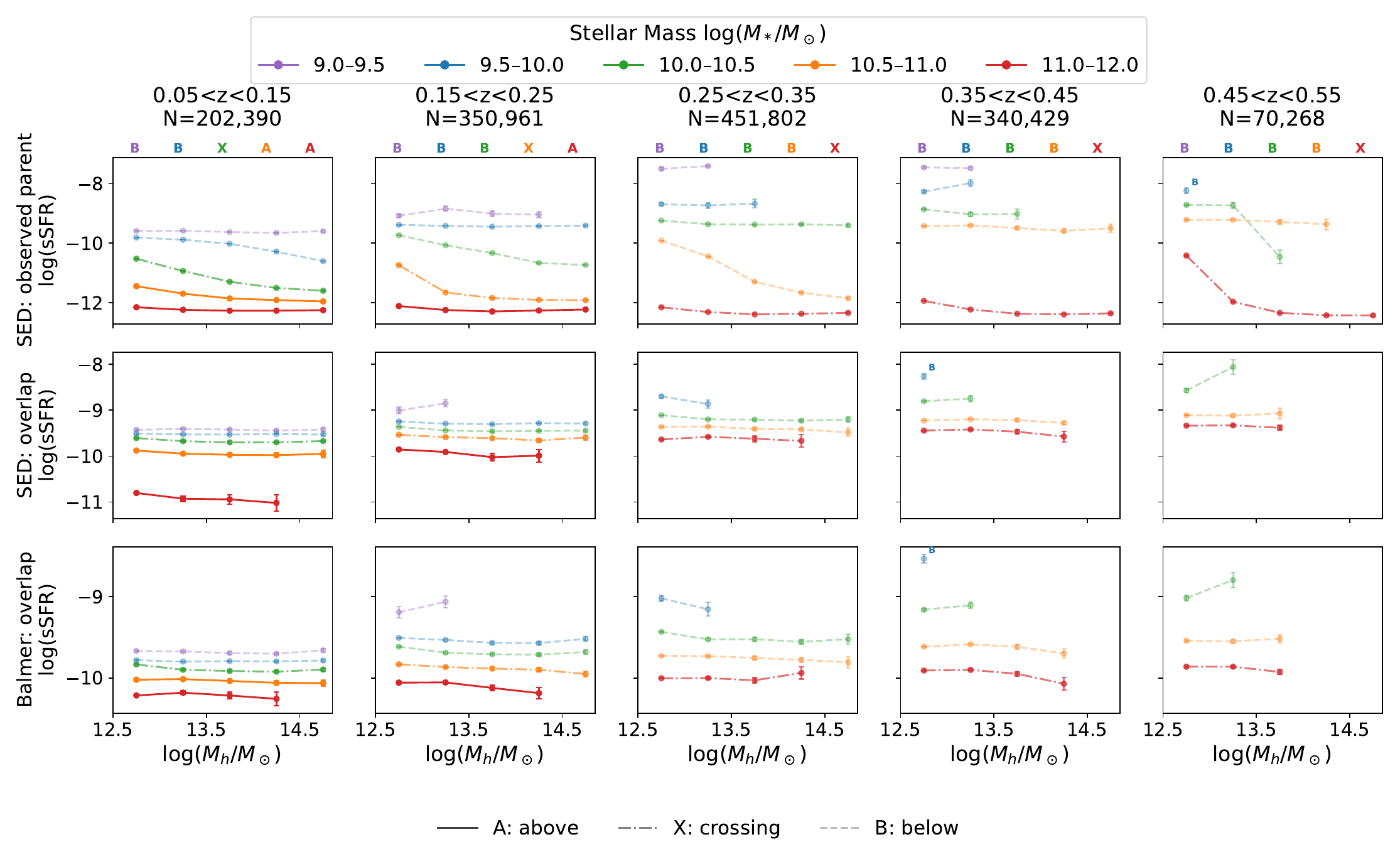}
\caption{Parent sSFR versus halo mass, showing selection and estimator effects separately. Row~1 shows the parent SED sSFR. Row~2 shows the SED sSFR for galaxies with Balmer estimates. Row~3 shows the aperture-corrected Balmer sSFR on a Chabrier scale for those same galaxies. Each measurement requires $N\geq30$; error bars are $1.253\,\mathrm{SD}/\sqrt{N}$, using the population standard deviation. Line styles and opacity follow Fig.~\ref{fig:distribution}. Comparisons in the text apply the completeness cut to the same galaxies.}
\label{fig:appendix_balmer_bridge}
\end{figure*}

Figure~\ref{fig:appendix_balmer_bridge} shows the parent comparison. Selected comparisons, summarised in Sect.~\ref{sec:sfrcompare}, apply the completeness cut to the same samples.
The transition Row~1$\to$2 isolates the sample-selection effect (same SED estimator, different galaxy subset), while Row~2$\to$3 isolates the estimator effect (same subset, SED versus Balmer).

The Balmer requirements preferentially remove quiescent galaxies from dense environments, raising sSFR in the common sample and generally weakening apparent cluster suppression. This sample has a higher overall satellite fraction, although changes in individual bins can have either sign.
Figure~\ref{fig:appendix_balmer_1to1_contour} compares SED and Balmer SFRs for parent galaxies measured by both methods.

\end{appendix}

\end{document}